\documentclass[superscriptaddress,pr,twocolumn,notitlepage,longbibliography]{revtex4-1}
\usepackage[english]{babel}
\usepackage{amsmath}
\usepackage{bm}
\usepackage{amsfonts}
\usepackage{amssymb}
\usepackage{makeidx}
\usepackage[colorlinks]{hyperref}
\usepackage[pdftex]{graphicx}
\usepackage{color}

\graphicspath{{img/}}

\newcommand{\Tr}{\mathop{\rm Tr}\nolimits}

\newcommand{\e}{\mathrm{e}}

\let\ifr\i
\renewcommand{\i}{{\rm i}}
\renewcommand{\d}{\mathrm d}
\renewcommand{\u}{\uparrow}
\renewcommand{\d}{\downarrow}
\newcommand{\U}{\Uparrow}
\newcommand{\D}{\Downarrow}
\definecolor{darkgreen}{rgb}{0.00, 0.75, 0.00}

\newcommand{\ket}[1]{\left|#1\right\rangle}

\begin{document}


\title{Measurement back-action and spin noise spectroscopy in a charged cavity-QED device in the strong coupling regime
}

\date{\today, file = \jobname.tex}

\author{D. S. Smirnov}
\affiliation{Ioffe Institute, 194021, St.-Petersburg, Russia}
\author{B. Reznychenko}
\author{A. Auff\`eves}
\affiliation{CEA/CNRS/UJF joint team ``Nanophysics and Semiconductors'', Institut N\'eel-CNRS, BP 166, 25 rue des Martyrs, 38042 Grenoble Cedex 9, France, Universit\'e Grenoble-Alpes \& CNRS, Institut N\`eel, Grenoble, 38000, France}
\author{L. Lanco}
\affiliation{Centre de Nanosciences et de Nanotechnologies, CNRS, Univ. Paris-Sud, Universit\'e Paris-Saclay, C2N -- Marcoussis, 91460 Marcoussis, France}
\affiliation{Universit\'e Paris Diderot, Sorbonne Paris Cit\'e, 75013 Paris, France}

\begin{abstract}
We study theoretically the spin-induced and photon-induced fluctuations of optical signals from a singly-charged quantum dot-microcavity structure. We identify the respective contributions of the photon-polariton interactions, in the strong light-matter coupling regime, and of the quantum back-action induced by photon detection on the spin system. Strong spin projection by a single photon is shown to be achievable, allowing the initialization and measurement of a fully-polarized Larmor precession. The spectrum of second-order correlations is deduced, displaying information on both spin and quantum dot-cavity dynamics. The presented theory thus bridges the gap between the fields of spin noise spectroscopy and quantum optics.
\end{abstract}

\maketitle

\section{Introduction}

A massive effort has been devoted to the development of quantum operations with single spins, used as stationary qubits which can be optically addressed. A number of systems have been proposed, such as single donor impurities in diamond~\cite{Maurer2012} or silicon~\cite{Steger2012}, or electron spins in semiconductor quantum dots (QDs)~\cite{Greilich2006}. A number of pioneering results have been obtained, for example, with single electron or hole spins in charged InAs/GaAs QDs: spin initialization~\cite{Atatuere2006,Xu2007,Gerardot2008} and read-out~\cite{Atatuere2007,Berezovsky2006,Mikkelsen2007}, as well as coherent spin manipulation~\cite{Press2008,Berezovsky2008}, spin-photon entanglement~\cite{Gao2012,DeGreve2012}, and spin-spin entanglement~\cite{Delteil2015}. 

In this general framework of quantum optics with single spins, a major goal is to increase the light-matter interaction efficiency. Optical spin read-out, for example, is usually based on a spin-dependent polarization rotation which is intrinsically small in the absence of cavity enhancement~\cite{Atatuere2007}. Using a charged quantum dot coupled to a microcavity, forming a cavity quantum electrodynamics (QED) system, this polarization rotation can be enhanced by several orders of magnitude~\cite{Hu2008a,Arnold2015,singleSpin}, and could potentially be used to develop a deterministic multiphoton entangler~\cite{Hu2008}. Ultrafast spin initialization can also be pushed down to a few tens of picoseconds, using accelerated spontaneous emission in a cavity-QED device~\cite{Loo2011}. The success rate of spin-photon or spin-spin entanglement experiments, currently limited by the low brightness of the QD emission~\cite{Delteil2015}, could also be drastically improved with cavity-enhanced emitters~\cite{Ding2016,Somaschi2016}. Finally, a spin-cavity interface could be used to implement quantum logic operations between a stationary spin qubit and a flying photonic qubit, as recently demonstrated with photonic crystals~\cite{Sun2016}. 

In parallel, in the framework of spin dynamics in semiconductors, a new experimental tool has emerged during the last decade: spin noise spectroscopy~\cite{Zapasskii:13,SinitsynReview}. It has already been successfully applied to bulk semiconductors~\cite{Oestreich_noise}, quantum wells~\cite{noise-trions}, and localized electrons~\cite{ZnOnoise,crooker2010}. The advances of this technique are conjugated with the increase of its sensitivity, in particular, due to the light matter interaction enhancement in microcavity structures~\cite{singleHole,NuclearNoise}. However the theoretical description of spin noise was previously developed only for the weak light-matter coupling regime~\cite{Glazov2016}, when the formation of polariton states does not take place.

At the interface between spin dynamics and quantum optics, we develop in this paper the theory of spin noise measurement and spectroscopy in the strong coupling regime using second-order photon correlations. We extend the spin noise spectroscopy technique to the ultimate limit of single spin dynamics, for a charged quantum dot within a cavity-QED device. We demonstrate that second-order photon correlations are not only governed by photon-polariton interactions, in the strong-coupling regime of cavity-QED, but also by the quantum back-action induced by a detected photon on the spin system. We provide a comprehensive analytical derivation of spin-induced photon correlations, in full agreement with numerical calculations. In particular, we show that perfect back-action by a single photon can be achieved to fully polarize a spin Larmor precession. We finally demonstrate that the spectrum of second-order correlations displays all the relevant information regarding spin dynamics and photon-polariton dynamics, and regarding the strength of the back-action induced by photon detection events.

The paper is organized as follows. In Sec.~\ref{sec:model} we present a theoretical model of a charged quantum dot microcavity under coherent driving. In Sec.~\ref{sec:timescales} we separate the short, intermediate and long timescales of spin and photon dynamics, and separately consider each of them to derive and analyze the evolution of second-order photon correlations. The analysis of the corresponding noise spectrum is presented in Sec.~\ref{sec:spectrum}, and the results are summarized in Sec.~\ref{sec:conclusion}.

\section{Model}

\label{sec:model}

\begin{figure}[t]
  \centering
  \includegraphics[width=\linewidth]{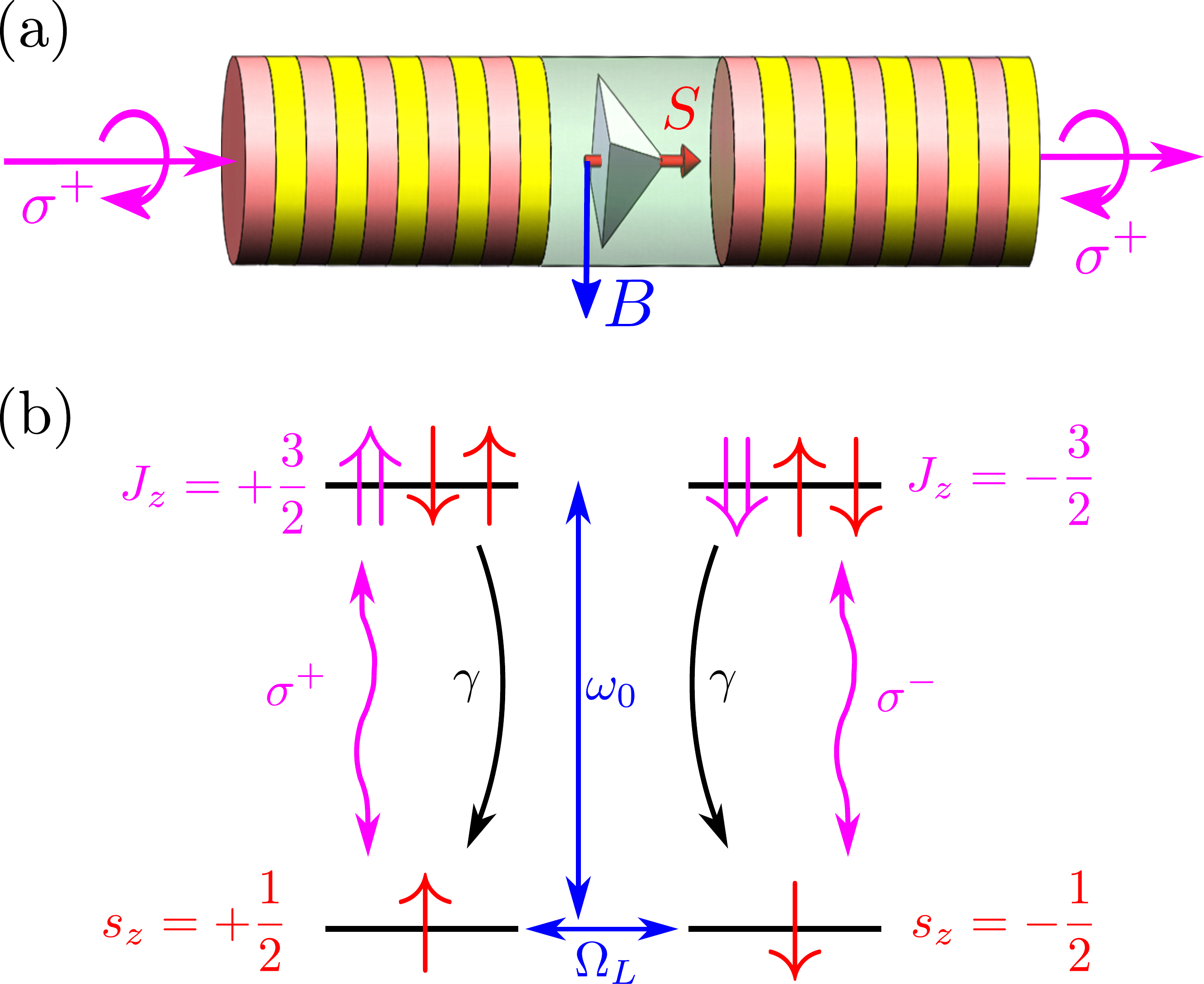}
  \caption{(a) Sketch of the quantum microcavity with an embedded quantum dot singly charged with an electron.
    (b)~Scheme of transitions between the quantum dot states. The optical transitions active in $\sigma^+$ and $\sigma^-$ polarizations are shown by wavy magenta arrows. The trion decay is shown by the black arrows, and the horizontal blue arrow denotes the Larmor precession of electron spin in the ground state.}
  \label{fig:sketch}
\end{figure}

We consider a zero-dimensional microcavity with an embedded quantum dot (QD) charged with an electron. The device and experiment under study are sketched in Fig.~\ref{fig:sketch}(a). The charged QD can be described as a four-level system with two ground states and two excited states, and with specific optical selection rules, as illustrated in Fig.~\ref{fig:sketch}(b) and as we now discuss. The Hamiltonian of the system in the presence of coherent excitation by the incident light can be presented as
\begin{multline}
 \label{ham}
 \mathcal H=\sum_{\pm}\left[\hbar\omega_c c_\pm^\dag c_\pm + \hbar\omega_0 a_{\pm 3/2}^\dag a_{\pm 3/2} + \frac{\hbar\Omega_L}{2}a^\dag_{\pm1/2}a_{\mp1/2}
\right.\\\left. +
 \left( \hbar gc_\pm^\dag a_{\pm 1/2}^\dag a_{\pm 3/2}
+ \hbar\mathcal{E}_\pm e^{-i\omega t} c_\pm^\dag + {\rm h.c.} \right)\right]\:.
\end{multline}
Here the cavity modes are assumed to be degenerate with respect to light polarization, the corresponding eigenfrequency is denoted as $\omega_c$, and $c_\pm$ ($c_\pm^\dag$) are the annihilation (creation) operators of the $\sigma^\pm$ photons in the cavity. We also assume the cavity resonance frequency to be close to the trion resonance frequency of the QD, $\omega_0$. The QD ground state is two fold degenerate with respect to the electron spin projection $s_z=\pm1/2$ on the growth direction, $z$. The excited trion state is formed by a pair of electrons in the singlet spin state and a heavy hole with spin projection $J_z=\pm3/2$. The annihilation and creation operators of the corresponding QD states are denoted as $a_{s_z}, a_{J_z}$ and $a_{s_z}^\dag, a_{J_z}^\dag$.  The last term in the first line in Eq.~\eqref{ham} stands for the Zeeman splitting of the QD ground state levels by the optional external magnetic field applied in Voigt geometry, with $\Omega_L$ the Larmor precession frequency. The spin precession in the trion state is neglected because of the small value of the transverse heavy hole $g$-factor~\cite{Mar99}. Furthermore, $g$ is the coupling strength and due to optical selection rules the photons are coupled only to the transitions between states with the same helicity, as shown in Fig.~\ref{fig:sketch}(b). Finally $\mathcal E_\pm$ are proportional to the circular components of the electric field in the incident electromagnetic field~\cite{milburn,singleSpin}.

The incoherent processes such as trion decay, photon escape from the cavity and spin relaxation should be described in the density matrix formalism. The quantum master equation for the density matrix $\rho(t)$ reads
\begin{equation}
\label{density:m}
  \dot\rho(t)=i[\rho(t),\mathcal H]-\mathcal L\lbrace\rho(t)\rbrace\:,
\end{equation}
where the Lindblad superoperator has the form~\cite{Poddubny2010,milburn,Carmichael}
\begin{align}
\label{Lindblad}
\mathcal L\lbrace\rho(t)\rbrace = \sum_i \gamma_i \left(
{\mathcal O_i}^\dag {\mathcal O_i}\rho(t)+\rho(t) {\mathcal O_i}^\dag {\mathcal O_i}-2{\mathcal O_i}\rho(t) {\mathcal O_i}^\dag
\right)
\end{align}
and operators $\mathcal O_i$ are taken from the set
$\left\{ c_\pm ,\, a_{\pm\frac{1}{2}}^\dag a_{\pm\frac{3}{2}},\, a_{\pm\frac{1}{2}}^\dag a_{\mp \frac{1}{2}},\, a_{+\frac{1}{2}}^\dag a_{+ \frac{1}{2}}-a_{-\frac{1}{2}}^\dag a_{-\frac{1}{2}}\right\}$
with corresponding decay rates 
$\left\{ {\varkappa} ,\, {\gamma},\, 1/\tau_s,\, 2/\tau_s  \right\}$. These rates are associated to cavity decay at rate $\varkappa$, trion decay unrelated with photon emission in the cavity mode at rate $\gamma$, and isotropic electron spin relaxation at rate $1/\tau_s$,
in the absence of pure dephasing processes. We abstain from describing microscopically the spin relaxation, which in moderate magnetic fields is usually dominated by the hyperfine interaction~\cite{merkulov02}. 

The general formalism of Eqs.~\eqref{ham}, \eqref{density:m} and \eqref{Lindblad} is valid for arbitrary relations  between the decay rates $\varkappa$, $\gamma$ and the photon-trion coupling constant $g$, therefore it is applicable in both the weak and strong coupling regimes.

In this paper we focus on the very strong coupling regime, when $g\gg\varkappa,\gamma$. For the sake of simplicity we also assume the perfect tuning between trion and cavity resonance frequencies, $\omega_c=\omega_0$, and we only consider $\sigma^+$ polarized-light incoming on the system, as described in Fig.~\ref{fig:sketch}(a). The first eigenstates of the system, excited in this case, are shown in Fig.~\ref{fig:levels}. The levels on the right, denoted as $\ket{\d,m\sigma^+}$, represent the tensorial product between the spin-down electron state $\left|\d\right\rangle$, corresponding to $s_z=-1/2$, and the Fock state $\left|m\sigma^+\right\rangle$, with $m=0,1,2,\ldots$ $\sigma^+$-polarized photons. Indeed, $\sigma^+$ light does not interact with this electron spin state due to the optical selection rules, see Fig.~\ref{fig:sketch}(b). By contrast the excited states for spin-up electron, shown on the left, represent the polariton states, that are the combinations of the trion and electron states with an equal number of quanta in the system~\cite{microcavities}:
\begin{equation}
  \frac{\left|\u,m\sigma^+\right\rangle\pm\ket{\U\d\u,(m-1)\sigma^+}}{\sqrt{2}},
  \label{eq:polaritons}
\end{equation}
where $\left|\u\right\rangle$ denotes electron spin-up $\left(s_z=+1/2\right)$ state, and $\left|\U\d\u\right\rangle$ stands for the trion $J_z=+3/2$ state. 
As can also be seen in Fig.~\ref{fig:levels}, no $\sigma^-$ polarized photons are involved. Indeed, we neglect the trion spin relaxation so that the $\ket{\D\u\d}$ trion state is not excited, therefore the system never emits $\sigma^-$ polarized photons, and the light transmitted or reflected by the system is always $\sigma^+$ polarized.

We study the transmission of $\sigma^+$ polarized light through the QD microcavity system, as shown in Fig.~\ref{fig:sketch}(a). We note that experimentally, it is easier to study the light reflected from the system, which consists of two contributions: the light reflected from the top Bragg mirror and the light re-extracted from the cavity. By contrast the transmitted light is related only to the latter contribution. Hence in order to  simplify the qualitative analysis and underline the cavity photon statistics we focus on the transmission geometry.

In thermal equilibrium the electron spin is unpolarized and the average transmission coefficient is $T_0$. However at every specific moment the electron is either in spin-up or in spin-down state, which means that the transmission coefficient, $T(t)$, unavoidably fluctuates in time. These fluctuations can be characterized by the correlator~\cite{Oestreich-review,Zapasskii:13}
\begin{equation}
  C(\tau)\equiv\frac{\left\langle\delta T(t)\delta T(t+\tau)\right\rangle}{T_0^2},
\label{eq:C}
\end{equation}
where $\delta T(t)=T(t)-T_0$ and the angular brackets denote averaging over the time $t$ for the given delay $\tau$. In accordance with the general properties of the correlators the correlation function is symmetric: $C(\tau)=C(-\tau)$~\cite{ll5_eng}. 

The transmitted field amplitude is proportional to the electric field amplitude inside the cavity, which is the basis of the input-output formalism~\cite{milburn}. As a result the second order photon correlation function 
\begin{equation}
  \label{eq:g2_def}
g^{(2)}(\tau)=
\frac{\left\langle c^\dag(t)c^\dag(t+\tau)c(t+\tau)c(t)\right\rangle}{\left\langle c^\dag(t) c(t)\right\rangle^2}
\end{equation}
is the same for the transmitted light and the light inside the cavity. Because we deal only with $\sigma^+$ polarized photons, we hereafter omit the subscripts ``$+$'' to shorten the notations unless it can lead to a confusion: $c$ thus stands for $c_+$ cavity photon annihilation operator. Taking into account normal ordering of $c_+$ and $c_+^\dag$ operators~\cite{milburn, Carmichael} it is straightforward to show that
\begin{equation}
  C(\tau)=g^{(2)}(\tau)-1,
\label{eq:Cg2}
\end{equation}
Therefore the transmittance noise measurement reveals, in addition to electron spin dynamics, photon statistics.

\section{Separation of timescales}

\label{sec:timescales}

Usually the spin dynamics takes place on the timescales longer, than the inverse decay rates of the excited states and period of the Rabi oscillations:
\begin{equation}
  \tau_s,\Omega_L^{-1}\ll\varkappa^{-1},\gamma^{-1},g^{-1}.
\end{equation}
Therefore it is possible to separate long timescales, when the electron spin dynamics takes place, and short timescales, when the photon-photon interactions play the dominant role.

\subsection{Short timescales}

On the short timescales, $t\ll\tau_s,\Omega_L^{-1}$, one can consider the two possible electron spin states independently.

\begin{figure}[t]
  \centering
  \includegraphics[width=\linewidth]{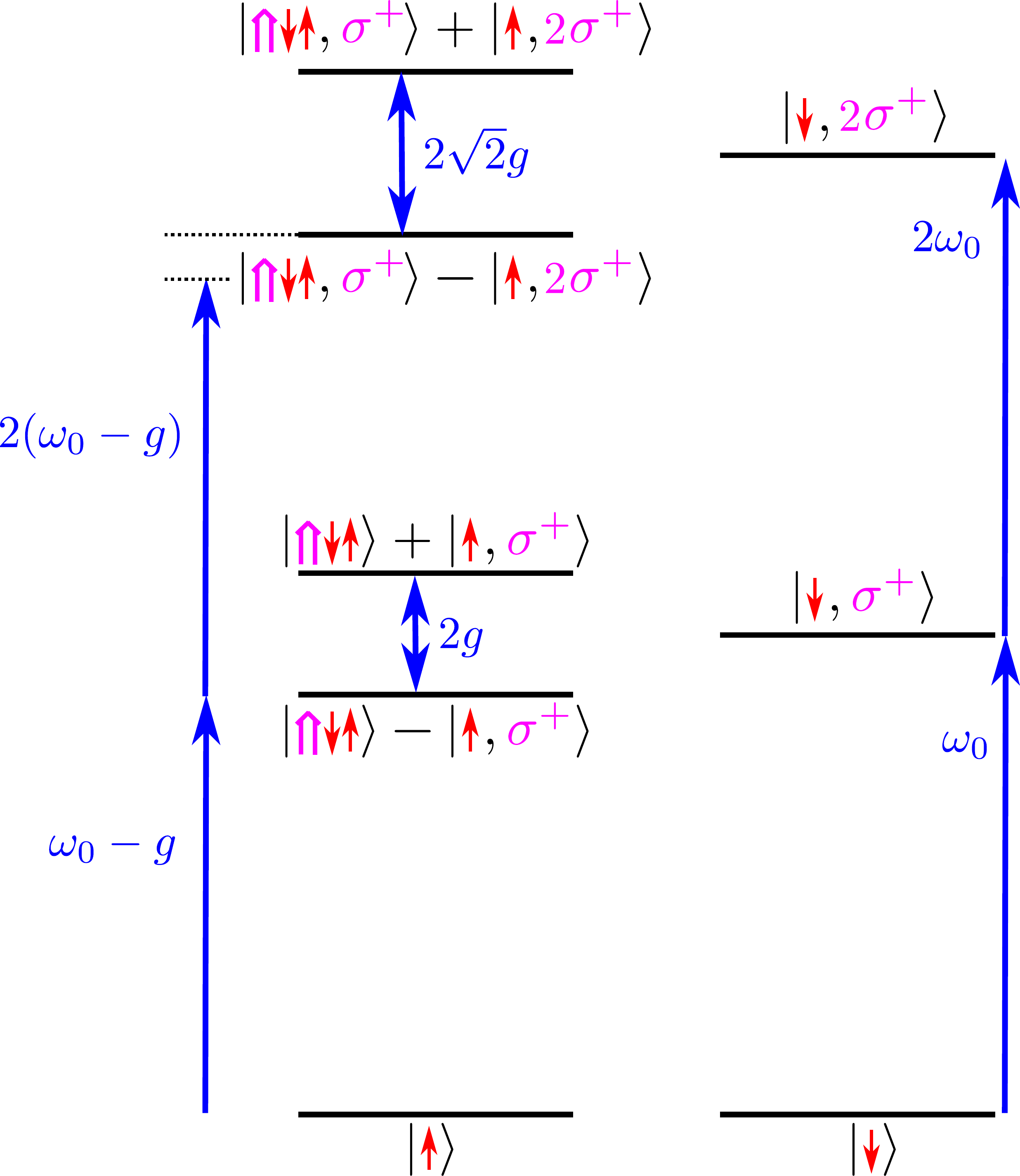}
  \caption{The scheme of the excited states and the splittings between them. The red arrows in the notations of the eigenstates denote electron spin, magenta arrows stand for heavy hole spins, and the photons are denoted as $\sigma^+$.}
  \label{fig:levels}
\end{figure}

\begin{figure}
  \centering
  \includegraphics[width=0.9\linewidth]{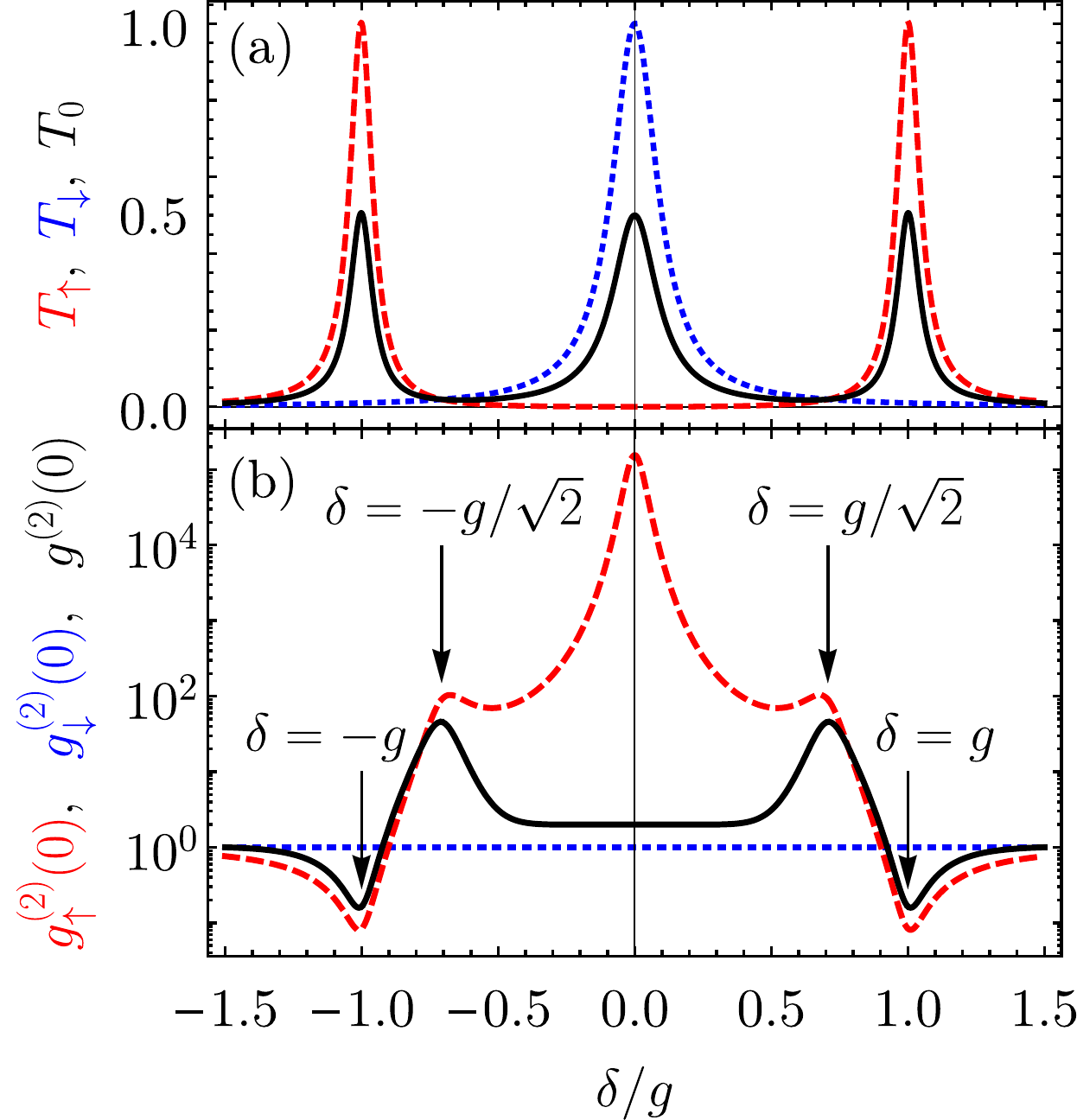}\\
  \caption{(a) The detuning dependence of the transmission coefficients, $T_\u$ and $T_\d$, red dashed and blue dotted lines, respectively, calculated after Eqs.~\eqref{eq:Td} and~\eqref{eq:Tu}. The solid black curve shows the spin-averaged transmission coefficient, $T_0$, Eq.~\eqref{eq:T0}. (b) The second order $\sigma^+$ photon correlation functions at zero delay time for spin-up and spin-down electron states, $g_{\u,\d}^{(2)}(0)$, red dashed and blue dotted lines, respectively, and $g^{(2)}(0)$ averaged over the spin states (black solid line) calculated after Eqs.~\eqref{eq:g2d}, \eqref{eq:g20u}, and~\eqref{eq:g2_short}. The parameters of the calculation are $g=10\varkappa$, $\gamma=0$, and $\omega_0=\omega_c$.
The arrows indicate the detunings $\delta$, corresponding to the resonant excitation of the four lowest polariton states.
}
  \label{fig:g20}
\end{figure}

In the case of spin-down electron, for the $\sigma^+$ polarized light under consideration the intracavity photons do not interact with the quantum dot, as discussed above and shown in Fig.~\ref{fig:levels}. Therefore the system effectively behaves as an empty cavity. The amplitude transmission coefficient in this case is
\begin{equation}
  \label{eq:td}
  t_\d=\frac{\i\varkappa}{\omega-\omega_c+\i\varkappa},
\end{equation}
where we have assumed that the cavity is symmetric and the two Bragg mirrors are characterized by equal transmission coefficients. The intensity transmission coefficient can be found as
\begin{equation}
  \label{eq:Td}
  T_\d=|t_\d|^2.
\end{equation}
It has a usual Lorentzian line shape centered at frequency $\omega=\omega_c$, as shown by the blue dotted line in Fig.~\ref{fig:g20}(a). The light in the cavity in this case is always coherent, and therefore
  \begin{equation}
    g^{(2)}_\d(\tau)=1,
    \label{eq:g2d}
  \end{equation}
where $g^{(2)}_\d$ stands for the second order photon correlation function, Eq.~\eqref{eq:g2_def}, in case of spin down electron.

In the case of spin-up electron, the polariton states are formed. For small light intensity, $\mathcal E\equiv\mathcal E_+ \ll \varkappa$, the system kinetics can be described by the Schroedinger-like equation~\cite{singleSpin}. We denote the probability amplitudes for the polariton states, Eq.~\eqref{eq:polaritons}, as $C_m^\pm$ ($m=1,2,\ldots$), respectively, and for the ground state $\left|\uparrow\right\rangle$ as $C_0$. The equation of motion for the first five states reads
\begin{subequations}
\label{eq:Shrodinger}
  \begin{equation}
    \dot{C}_0=0,
    \label{eq:Shrodinger0}
  \end{equation}
  \begin{equation}
    \i\dot{C}_1^\pm=(\omega_0\pm g-\i\varkappa_1)C_1^\pm+\frac{C_0}{\sqrt{2}}\mathcal E\e^{-\i\omega t},
    \label{eq:Shrodinger1}
  \end{equation}
  \begin{multline}
    \i \dot{C}_2^\pm=(2\omega_0\pm\sqrt{2}g-\i\varkappa_2)C_2^\pm\\+\left(\frac{\sqrt{2}+1}{2}C_1^\pm+\frac{\sqrt{2}-1}{2}C_1^\mp\right)\mathcal E\e^{-\i\omega t}.
    \label{eq:Shrodinger2}
  \end{multline}
\end{subequations}
Here the decay rate for the first doublet of polariton states, $m=1$, is calculated as an average of the photon and trion decay rates: $\varkappa_1=(\gamma+\varkappa)/2$, and, similarly, $\varkappa_2=(\varkappa+\gamma+2\varkappa)/2$ for the second doublet of polariton states. Though the presented formalism remains valid for arbitrary relation between $\gamma$ and $\varkappa$, below we consider the experimentally relevant so-called ``bad cavity'' case, when $\gamma\ll\varkappa\ll g$.  In this limit $\varkappa_1=\varkappa/2$ and $\varkappa_2=3\varkappa/2$.

From Eq.~\eqref{eq:Shrodinger0} one can see that $C_0(t)=\rm{const}$, and one can take $C_0(t)=1$. From Eqs.~\eqref{eq:Shrodinger1} and~\eqref{eq:Shrodinger2} one finds for the steady state
\begin{subequations}
  \label{eq:steady}
  \begin{equation}
    C_1^\pm=\frac{1}{\delta\mp g+\i\varkappa/2}\frac{\mathcal E}{\sqrt{2}}\e^{-\i\omega t},
    \label{eq:C1}
  \end{equation}
  \begin{equation}
    C_2^\pm=\frac{\left(\sqrt{2}+1\right)C_1^\pm+\left(\sqrt{2}-1\right)C_1^\mp}{2\delta\mp \sqrt{2}g+3\i\varkappa/2}\frac{\mathcal E}{2}\e^{-\i\omega t},
  \end{equation}
\end{subequations}
where we have introduced the detuning from the bare cavity mode $\delta=\omega-\omega_0$.
Note that $C_1^\pm\sim\mathcal E$ and $C_2^\pm\sim\mathcal E^2$, hence in the limit of low incident light intensity, the amplitude probabilities $C_1^\pm$ contain all the relevant information to compute the transmitted light intensity, while the amplitude probabilities $C_2^\pm$ contain all the relevant information to compute the second-order correlation statistics of the transmitted light.

In the low-power limit the amplitude of the transmitted wave is determined by $\left\langle c\right\rangle\approx\left(C_1^++C_1^-\right)C_0^*/\sqrt{2}$ and, since no pure dephasing is assumed, the transmitted light is fully coherent, $\left\langle c^\dag c\right\rangle=\left|\left\langle c\right\rangle\right|^2=\left|\left(C_1^++C_1^-\right)/\sqrt{2}\right|^2$. Hence the amplitude and intensity related transmission coefficients can be found, respectively, as~\cite{Hu2008a,englund2007controlling}
\begin{equation}
  t_\u=\frac{\i\varkappa(C_1^+ + C_1^-)}{\sqrt{2}\mathcal E\e^{-\i\omega t}}=\frac{\i\varkappa}{\delta+\i\varkappa-{g^2}/{\delta}},
  \label{eq:tu}
\end{equation}
\begin{equation}
  T_\u=\frac{\varkappa^2\left|\left(C_1^++C_1^-\right)\right|^2}{2\mathcal E^2}=|t_\u|^2,
  \label{eq:Tu}
\end{equation}
where we have taken into account that $g\gg\varkappa$. Eq.~\eqref{eq:Tu} is shown by the red dashed curve in Fig.~\ref{fig:g20}(a). It describes a pair of Lorentzian peaks, centered at polariton frequencies, $\delta=\pm g$, with the width $\varkappa/2$.

Similarly, from the equation $\left\langle c^\dag c^\dag cc\right\rangle=\left|C_2^++C_2^-\right|^2$ one obtains the expression for the second order correlator
\begin{equation}
  g^{(2)}_\u(0)=4\frac{\left|C_2^++C_2^-\right|^2}{\left|C_1^++C_1^-\right|^4}.
\label{eq:g20u}
\end{equation}
This expression is shown in Fig.~\ref{fig:g20}(b) by the red dashed line as a function of the detuning. It reaches the maximum value at $\delta=0$ and has local maxima at $\delta=\pm g/\sqrt{2}$, which are indicated by the arrows. Moreover $g^{(2)}_\u(0)$ drops down at the detunings $\delta=\pm g$ and has asymmetric minima at these frequencies.

The dependence of $g^{(2)}_\u(0)$ on the frequency can be easily understood considering the scheme of energy levels in the system, shown in Fig.~\ref{fig:levels}. When the incident light is in resonance with one of the first two polariton states, $\delta=\pm g$, it efficiently excites the states with $m=1$. However the higher polariton states are weakly populated, because of the different splitting between those states, see the blue arrows on the left in Fig.~\ref{fig:levels}. This means that the system tends to behave as a two-level atom~\cite{carmichael1985photon,poshakinskiy2014time}, which results into antibunching $g^{(2)}_\u(0)\sim(\varkappa/g)^2\ll 1$. By contrast, the detunings $\delta=\pm g/\sqrt{2}$ correspond to the efficient excitation of the second polariton doublet, giving rise to two-photon absorption, while the lower doublet is off-resonant. At the corresponding frequencies the light is strongly bunched, $g^{(2)}_\u(0)\sim(g/\varkappa)^2\gg1$. Finally at zero detuning all the polariton states are non-resonant with the incoming laser, and are thus very weakly excited, leading to a small transmitted light intensity. In addition, for the first polariton states $C_1^+\approx-C_1^-$, so 
 the transmitted light is very strongly bunched, $g^{(2)}_\u(0)\sim(g/\varkappa)^4$. However the observation of this effect is complicated by the weak transmittance in the spin up state: in the case of small detunings $T_\u\ll 1$ while $T_\d\approx1$.

We now turn to the time dependence of the second order correlation function, which can be found by solving the Schroedinger equation, Eq.~\eqref{eq:Shrodinger}, with the initial condition $\tilde\Psi(0)=c\Psi_0(0)$ at $\tau=0$, where $c$ is the photon annihilation operator and $\Psi_0(0)$ the steady state system wave function, Eq.~\eqref{eq:steady}, and $\tilde\Psi(\tau)$ describes the system evolution after a single photon escape from the cavity. In the following the `tilde' notation, as in $\tilde\Psi$, will be used for all the quantities computed after a single-photon detection event. In the basis of polariton states the initial conditions are
\begin{subequations}
  \begin{equation}
    \tilde C_0(0)=\frac{C_1^+(0)+C_1^-(0)}{\sqrt{2}},
    \label{eq:C00}
  \end{equation}
  \begin{equation}
    \tilde C_1^\pm(0)=\frac{\sqrt{2}+1}{2}C_2^\pm(0)+\frac{\sqrt{2}-1}{2}C_2^\mp(0).
  \end{equation}
\end{subequations}
The solution of the Schrodinger equation, Eq.~\eqref{eq:Shrodinger}, in the time domain with these initial conditions reads $\tilde C_0(\tau)=\tilde C_0(0)$ and
\begin{multline}
  \tilde C_1^\pm(\tau)=\tilde C_1^\pm(0)\e^{-\i(\omega_0\pm g)\tau}\e^{-\varkappa\tau/2}\\+\frac{\tilde C_0(0)/\sqrt{2}}{\omega-\omega_0\mp g+\i\varkappa/2}\left[1-\e^{\i(\delta\mp g)\tau}\e^{-\varkappa\tau/2} \right]\mathcal E\e^{-\i\omega\tau}.
\label{eq:C1tilde}
\end{multline}
Finally the second order photon correlator can be found as
\begin{equation}
  g^{(2)}_\u(\tau)=2\frac{\left|\tilde C_1^+(\tau)+\tilde C_1^-(\tau)\right|^2}{\left|C_1^++C_1^-\right|^4}.
\label{eq:g2u}
\end{equation}

\begin{figure}[t]
  \centering
  \includegraphics[width=\linewidth]{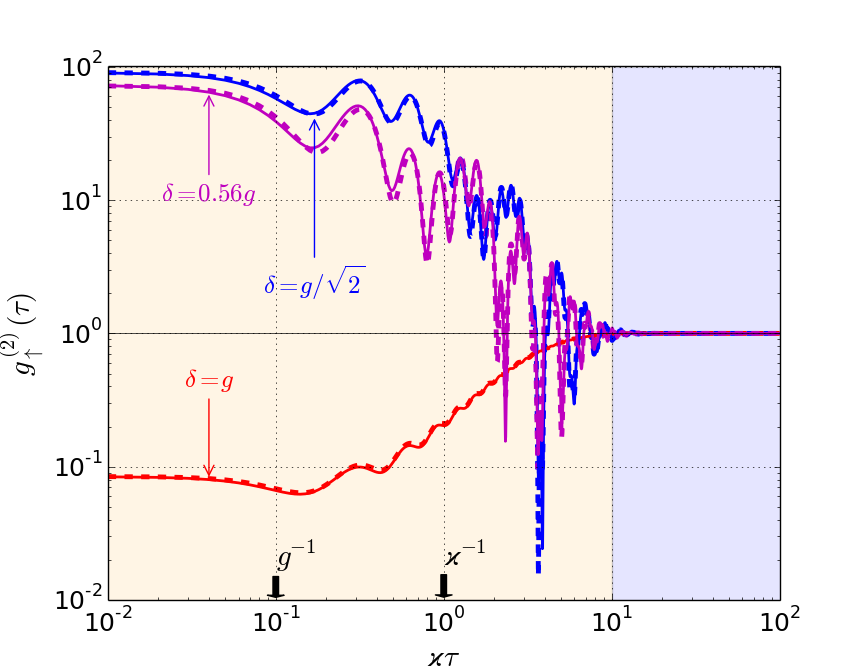}
  \caption{Time dependence of the second order correlator $g_\u^{(2)}(\tau)$ in log-log scale for different detunings. The parameters of calculation are the same as in Fig.~\ref{fig:g20}. The solid lines are calculated numerically using the full Hamiltonian~\eqref{ham} and Lindblad operators~\eqref{Lindblad}. The dashed lines represent analytical calculation after Eq.~\eqref{eq:g2u}. The background colors indicate the timescale separation: yellow for the short and light blue for the intermediate timescales, see the main text for details.
}
  \label{fig:g2t_u}
\end{figure}

The time dependence of $g^{(2)}_\u(\tau)$ is shown in Fig.~\ref{fig:g2t_u} for different detunings which will be commented later on. In this Figure the photon correlation function is computed with two approaches: appart from the described analytical approach of Eq.~\eqref{eq:g2u}, the photon correlation function can be calculated numerically using the Lindblad equation~\eqref{density:m}. 
For all detunings the agreement is perfect, and is limited only by the small parameter $\varkappa/g$ and the low incident light power $\mathcal E\ll\varkappa$.
In the general case the correlation function beats at frequencies $2g$ and $|\delta\pm g|$ and decays to unity. The oscillations start after detection of a photon because the wave function $\tilde\Psi$ is not stationary. The Rabi oscillations between the first polariton states give rise to the oscillations at frequency $2g$. Population of the first polariton states due to the excitation of the ground state after a photon detection also results in oscillations at frequencies $|\delta\pm g|$ corresponding to the Rabi oscillations between the ground and polariton states, as described by Eq.~\eqref{eq:C1tilde}. At time delays $\tau\gg\varkappa^{-1}$ one has $\left|\tilde C_1^+(\tau)+\tilde C_1^-(\tau)\right|=\left|C_1^++C_1^-\right|^2/\sqrt{2}$, and from Eq.~\eqref{eq:g2u} one can see that the beatings decay and $g^{(2)}_\u(\infty)=1$, as expected.

\subsection{Averaging over spin states}


In the previous subsection we have calculated separately the second order correlation functions $g^{(2)}_\u(\tau)$ and $g^{(2)}_\d(\tau)$, considering the cases of a fixed spin-up and spin-down states. However the unpolarized electron spin state is an incoherent superposition of two states $\left|\u\right>$ and $\left|\d\right>$. The density matrix of the system in this case is a superposition of the steady state density matrices corresponding to the two possible electron spin orientations: $\rho=(\rho_\u+\rho_\d)/2$. For any operator $O$ the quantum mechanical average can be calculated as
\begin{equation}
  \left<O\right>=\Tr(\rho O)=\frac{\Tr(\rho_\u O)+\Tr(\rho_\d O)}{2}\equiv\frac{\left<O\right>_\u+\left<O\right>_\d}{2},
\label{eq:average}
\end{equation}
where we have introduced the notations $\left<O\right>_{\u/\d}$ for the operator expectation values in electron spin-up/down steady state.

From Eq.~\eqref{eq:average} for $O=c^\dag c$ one finds the transmission coefficient in the case of unpolarized electron spin
\begin{equation}
  \label{eq:T0}
  T_0=\frac{T_\u+T_\d}{2}.
\end{equation}
It is shown in Fig.~\ref{fig:g20}(a) by the black solid curve as a function of the detuning, and is characterized by three peaks with the amplitude $1/2$. Note that due to the averaging over spin states the transmitted light is not fully coherent, therefore $T_0\neq|t|^2$, where $t=(t_\u+t_\d)/2$ is the averaged amplitude transmission coefficient.

\begin{figure}[t]
  \centering
  \includegraphics[width=\linewidth]{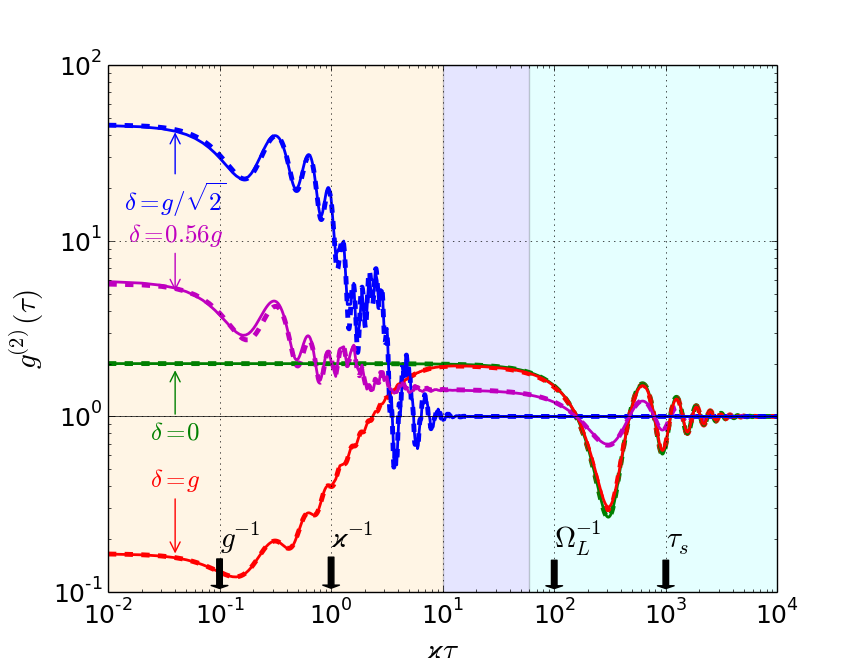}
  \caption{Time dependence of the spin-averaged second order correlator $g^{(2)}(\tau)$ for different detunings. The parameters of calculations are the same as in Fig.~\ref{fig:g20} and $\Omega_L/\varkappa=0.01$, $\tau_s\varkappa=1000$. The solid lines, as in Fig.~\ref{fig:g2t_u}, are calculated numerically using the full Hamiltonian~\eqref{ham} and Lindblad operators~\eqref{Lindblad}. The dashed lines represent analytical calculation after Eq.~\eqref{eq:Call}. The background colors indicate the timescale separation: yellow for the short ones, light blue for the intermediate and light green for the long timescales.}
  \label{fig:g2t}
\end{figure}

Applying Eq.~\eqref{eq:average} to the operator $c^\dag(0)c^\dag(\tau)c(\tau)c(0)$ we arrive at the expression for the spin-averaged second order correlator
\begin{equation}
  \label{eq:g2_short}
    g^{(2)}(\tau)=2\frac{T_\u^2g^{(2)}_\u(\tau)+T_\d^2}{\left(T_\u+T_\d\right)^2},
\end{equation}
where we have taken into account Eq.~\eqref{eq:g2d}. As will be shown in the next section, Eq.~\eqref{eq:g2_short} can be given an interpretation in terms of the probabilities that the spin is in the up or down state after a single photon detection.

The dependence of $g^{(2)}(0)$ on the detuning is shown in Fig.~\ref{fig:g20}(b) by the black line. It is similar to $g^{(2)}_\uparrow(0)$ with the exception for the region of small detunings, $\left|\delta\right| \ll g$. Here $T_\uparrow\ll T_\downarrow$ and Eq.~\eqref{eq:g2_short} yields $g^{(2)}(0)\approx 2$.

Eq.~\eqref{eq:g2_short} can be used to compute the full time dependance of the second order correlation $g^{(2)}(\tau)$, at short and intermediate timescales, for various detunings. The result of the calculation is shown in Fig.~\ref{fig:g2t} for the same detunings as in Fig.~\ref{fig:g2t_u} along with the calculation for long timescales which will be commented later on. Here also the photon correlation function has been computed with the analytical approach of Eq.~\eqref{eq:g2_short} and calculated numerically with the Lindblad equation~\eqref{density:m}; the agreement is perfect in the limit $\varkappa \ll g$ and under a very low excitation $\mathcal E\ll\varkappa$. At short timescale, we find that the time dependance of $g^{(2)}(\tau)$ is dominated by the photon dynamics in the cavity and is essentially given by $g^{(2)}_\u(\tau)$, shown in Fig.~\ref{fig:g2t_u}, with a change of scale induced by the spin averaging of Eq.~\eqref{eq:g2_short}. At longer timescales, as discussed below, further analysis is required to understand the characteristics of the full photon correlation dynamics.


\begin{figure}
  \centering
  \includegraphics[width=0.9\linewidth]{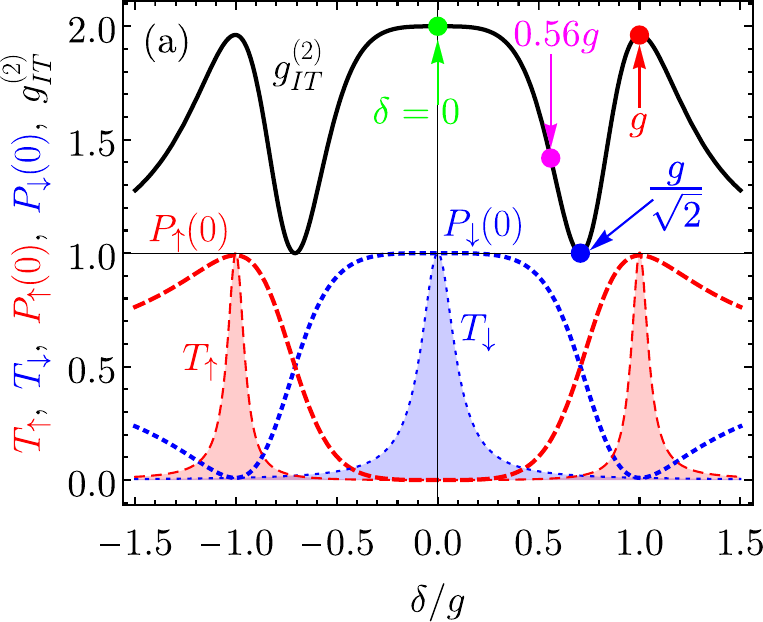}\\
  \includegraphics[width=\linewidth]{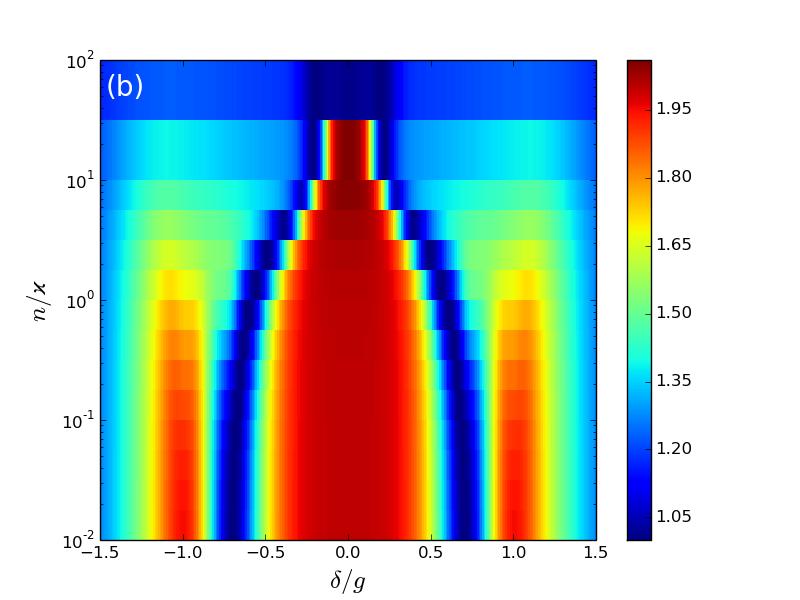} 
  \caption{(a) The frequency dependence of $g^{(2)}_{IT}$ (black solid line) calculated after Eq.~\eqref{eq:g2med}. The thick red dashed and thick blue dotted lines show $\tilde{P}_\u$ and $\tilde{P}_\d$, calculated after Eq.~\eqref{eq:Pud}, respectively. The thin red dashed and thin blue dotted lines with the filling show the transmission coefficients, $T_{\u,\d}$, calculated after Eqs.~\eqref{eq:Td} and~\eqref{eq:Tu}. The dots on the black curve denote the detunings used for the calculations in Figs.~\ref{fig:g2t_u} and~\ref{fig:g2t}. 
  (b) Numerical simulation of $g^{(2)}_{IT}$ dependence on $\delta$ and on the average number of photons incident on the cavity per photon lifetime $n$.
The parameters of the calculations are the same as in Fig.~\ref{fig:g20}.
}
  \label{fig:g2med}
\end{figure}

\subsection{Intermediate timescales}

For moderately long delays $\tau$, such as $\varkappa^{-1}\ll\tau\ll\tau_s,\Omega_L^{-1}$ one has $g_\u^{(2)}(\tau)\to1$, see Fig.~\ref{fig:g2t_u}. Therefore the spin-averaged second order correlator is constant, $g^{(2)}(\tau)\equiv g^{(2)}_{IT}$, where ``IT'' stands for ``intermediate timescale''. 
From Eq.~\eqref{eq:g2_short} we find:
\begin{equation}
    g^{(2)}_{IT}=1+\left(\frac{T_\u-T_\d}{T_\u+T_\d}\right)^2.
\label{eq:g2med}
\end{equation}
Clearly this expression takes values within the range $[1;2]$. One can see, that indeed in Fig.~\ref{fig:g2t}, at intermediate timescales $10<\tau\varkappa<100$, $g^{(2)}(\tau)$ is constant.

The detuning dependence of $g^{(2)}_{IT}$ is shown in Fig.~\ref{fig:g2med}(a). It consists of three peaks, corresponding to the peaks in $T_0$. At the bare cavity resonance $\delta=0$ one has $T_\d\gg T_\u$, and therefore $g^{(2)}_{IT}\approx2$. The same situation takes place at $\delta=\pm g$, when $T_\u\gg T_\d$. By contrast at $\delta=\pm g/\sqrt{2}$ the transmission coefficients are equal, $T_\u=T_\d$, so $g^{(2)}_{IT}=1$. At large detunings, $\delta\gg g$ one can see that $g^{(2)}_{IT}$ also tends to unity. Finally, the detuning $\delta=0.56 g$ corresponds to a general intermediate situation where $T_\u \neq T_\d$ but where both transmission coefficients are non-negligible, leading to an intermediate value $1 < g^{(2)}_{IT} < 2$. All these results are also illustrated in Fig.~\ref{fig:g2t}, where at intermediate timescale $g^{(2)}$ takes different constant values between 1 and 2.

We now turn to the interpretation of Eq. \eqref{eq:g2med} in terms of spin-projective measurements. After single photon detection at time $t=0$ the density matrix of the QD-cavity system takes the form
  \begin{equation}
    \tilde \rho(0) = \frac{c_+\rho_{ss} c_+^\dag}{\Tr\left(c_+\rho_{ss} c_+^\dag\right)},
  \end{equation}
where $\rho_{ss}$ denotes the stationary density matrix. The post-detection density matrix $\tilde \rho$ is not stationary, and thus evolves with time towards $\rho_{ss}$. We introduce the probabilities to find an electron in spin up and down states determined by the operators
\begin{equation}
  \hat{P}_{\u,\d}=a_{\pm\frac{1}{2}}^\dag a_{\pm \frac{1}{2}},
\end{equation}
respectively. At short and intermediate timescales the electron spin dynamics can be neglected, which gives the expectation values
\begin{equation}
  \tilde{P}_{\u,\d}(0)=\Tr\left(\hat{P}_{\u,\d}\tilde\rho(0)\right)=\frac{T_{\u,\d}}{T_\u+T_\d}.
\label{eq:Pud}
\end{equation}
In the above expression $\tilde{P}_{\u,\d}$ denote the probabilities of the system to be in the up or down ground state after a first photon detection event. The photon correlation function at the intermediate timescale, given by Eq. \eqref{eq:g2med}, can then be presented as
\begin{equation}\label{eq:g2_sz}
  g^{(2)}_{IT}=1+4\tilde{S}_z^2(0),
\end{equation}
where $\tilde{S}_z=(\tilde{P}_\u-\tilde{P}_\d)/2$ is the average spin after a first photon detection. This expression shows that the photon correlations at intermediate timescales are directly linked to a spin polarization, obtained via the back-action induced by the photon detection event on the spin system.

The back-action of such a measurement onto the spin system can be understood through Eqs.~\eqref{eq:Pud} and \eqref{eq:g2_sz}. If $T_\u\ll T_\d$, it is a strong projective measurement, because $\tilde{P}_\d=1$, while $\tilde{P}_\u=0$, i.e. after the detection the spin is projected into $\ket{\d}$, leading to $\tilde S_z(0)=-1/2$ and $g^{(2)}_{IT}=2$. It is the case for small detunings $|\delta|\ll g$, when the transmitted light is in resonance with the empty cavity mode. By contrast, at the detunings $\delta\approx \pm g$ the polariton states are in resonance, which leads to $T_\u\gg T_\d$. In this case $\tilde{P}_\u=1$, while $\tilde{P}_\d=0$, thus $\tilde S_z(0)=+1/2$ and $g^{(2)}_{IT}=2$. It is also a strong projective measurement, where the spin is projected into $\left|\u\right\rangle$.
If the transmission coefficients are equal, $T_\u=T_\d$, then projection probabilities are equal as well, $\tilde{P}_\u=\tilde{P}_\d$, i.e. no backaction is observed and the spin stays unpolarized: $\tilde S_z(0)=0$ and $g^{(2)}_{IT}=1$. It is the case at large detunings, $|\delta|\gg g$, and also at $\delta=\pm g/\sqrt{2}$. In the intermediate situation $\delta=0.56g$ the spin is partially projected in the down state, as $T_\u < T_\d$ and thus $\tilde{P}_\u<\tilde{P}_\d$, leading to $-\frac{1}{2}<\tilde S_z(0)<0$. The probabilities $\tilde{P}_{\u,\d}$ are displayed in Fig.~\ref{fig:g2med} by thick red dashed and thick blue dotted lines, respectively.

Therefore, the fact that $g^{(2)}_{IT}=2$ is a signature of a strong projective measurement. Indeed, $g^{(2)}_{IT}$ can be understood as a value proportional to the conditional probability to detect a photon, if one photon has already been detected. In the $T_\d \ll T_\u$ case, even though the initial occupation probabilities of both spin orientations are equal, $P_\u=P_\d=0.5$, only the spin-up case can lead to a photon detection. The conditional probability of detecting a photon is thus multiplied by two after a first photon detection event, as this first detection event has increased the spin-up probability from $P_\u=0.5$ to $\tilde{P}_\u=1$. The same reasoning is valid for the case when $T_\u \ll T_\d$.

The expression Eq.~\eqref{eq:g2med} for the photon correlation function at intermediate timescales, can be also obtained making use of the relation with the correlator of the transmission coefficient, Eq.~\eqref{eq:Cg2}. Indeed at the intermediate timescales 
 the electron spin can be considered as frozen. Therefore the transmission coefficient at $t$ and $t+\tau$ is the same: either $T_\u$ or $T_\d$. In both cases the absolute value of the fluctuation is the same $\left|\delta T\right|=\left|T_\u-T_\d\right|/2$ and $\left\langle\delta T(t)\delta T(t+\tau)\right\rangle=\delta T^2$. Taking into account the average value of transmission coefficient, Eq.~\eqref{eq:T0}, one obtains
  \begin{equation}
    \mathcal C_{IT}=\left(\frac{T_\u-T_\d}{T_\u+T_\d}\right)^2
  \end{equation}
for $\mathcal C(\tau)$ at intermediate timescales. This expression corresponds to Eq.~\eqref{eq:g2med} for the second order photon correlation function.

We note that this analytical approach is limited to the low power limit, but direct numerical calculations can be performed in principle for any incoming power. To illustrate this we have numerically studied the modification of the $g^{(2)}_{IT}$ dependence on $\delta$ with the increase of light power. Fig.~\ref{fig:g2med}(b) shows this modification as a function of the average number of photons incident on the cavity per photon lifetime, $n=|\mathcal{E}|^2/\kappa$. With increasing power, the central peak in the detuning dependence becomes narrower, and the side peaks grow wider, as more polariton modes are excited. Both peaks are decreasing, so that above an incoming power greater than $n \sim g^2/\varkappa$, the QD transition is saturated. In such a case the system starts to behave as an empty cavity, i.e. $T_\u \approx T_\d$, hence the spin projection becomes inefficient: $\tilde S_z(0)\approx 0$ and $g^{(2)}_{IT}$ reaches unity for all values of $\delta$. It signifies that projective measurements are no longer possible at high power.

\subsection{Long timescales}



Finally we consider long delays $\tau$, of the order of Larmor precession period and spin relaxation time. At the long timescales the transmission coefficient adiabatically follows the electron spin state:
\begin{equation}
  \label{eq:Ttau}
  T(\tau)=T_\u \tilde{P}_\u(\tau) + T_\d \tilde{P}_\d(\tau).
\end{equation}
Indeed $\tilde{P}_\u(\tau)$ and $\tilde{P}_\d(\tau)$ slowly evolve compared to the other characteristic timescales ($g^{-1}$ and $\kappa^{-1}$): a quasiequilibrium is established at each delay, and the probability amplitudes of the polariton states are proportional to Eq.~\eqref{eq:C1}. Therefore the transmission coefficient is a linear superposition of $T_\u$ given by Eq.~\eqref{eq:Tu} and $T_\d$ given by Eq.~\eqref{eq:Td} with the corresponding probabilities $\tilde{P}_\u(\tau)$ and $\tilde{P}_\d(\tau)$. The photon-photon interaction plays no role at the long timescales.

From the Hamiltonian~\eqref{ham} and Lindblad operators~\eqref{Lindblad} one directly finds that, as usual, the spin-up and spin-down state probabilities oscillate at Larmor frequency and decay towards $1/2$ as
\begin{equation}
  \label{eq:Ptau}
  \tilde{P}_{\u,\d}(\tau)=\left(\tilde{P}_{\u,\d}(0)-\frac{1}{2}\right)\cos\left(\Omega_L \tau\right)\e^{-\tau/\tau_s}+\frac{1}{2}.
\end{equation}
Using the definition, Eq.~\eqref{eq:C}, and the relation Eq.~\eqref{eq:Cg2}, one obtains the standard expression for the long timescale correlator
\begin{equation}
  g^{(2)}(\tau)=1+4\tilde{S}_z(0)\tilde{S}_z(\tau),
  \label{eq:C_long}
\end{equation}
where
\begin{equation}
  \tilde{S}_z(\tau)=\tilde{S}_z(0)\cos\left(\Omega_L \tau\right)\e^{-\tau/\tau_s}.
  \label{eq:Stau}
\end{equation}
This is a very well known result in spin noise spectroscopy~\cite{Zapasskii:13,SinitsynReview}.
Explicitly the transmission coefficient correlator at long timescales can be related to the spin correlation function as
  \begin{equation}
    \mathcal C(\tau)=\mathcal 4 C_{IT}\left\langle{S_z(0)S_z(\tau)}\right\rangle,
  \end{equation}
where we have taken into account, that $\left\langle S_z^2\right\rangle=1/4$.

Eq.~\eqref{eq:C_long} describes the oscillations of the transmittance correlator at long timescales, which are shown in Fig.~\ref{fig:g2t}. For the calculation we have used $\Omega_L \tau_s=10$, though the presented formulas are valid for arbitrary relation between $\Omega_L$ and $\tau_s$. The oscillations begin at $\tau\gtrsim\Omega_L^{-1}$ and decay during the time $\tau_s$~\footnote{In the absence of external magnetic field $\Omega_L=0$, the spin relaxation leads to the monoexponential decay of the electron spin correlation function.}. The amplitude of oscillations depend on the detuning and reaches its maximum when the difference $|T_\u-T_\d|$ is the largest, i.e. at $\delta=0$ and~$\delta=\pm g$. On the contrary, at $\delta=\pm\frac{g}{\sqrt{2}}$ $T_\u=T_\d$ and the spin stays completely unpolarized: $\tilde{S}_z(\tau)=\tilde{S}_z(0)=0$. Finally, the case $\delta=0.56 g$ leads to oscillations with an intermediate contrast, due to the imperfect spin projection $-1/2<\tilde S_z(0)<0$. These long timescale correlations can be conveniently used for the experimental measurement of the spin relaxation time and effective $g$-factor in this system.   

\begin{figure}
  \includegraphics[width=\linewidth]{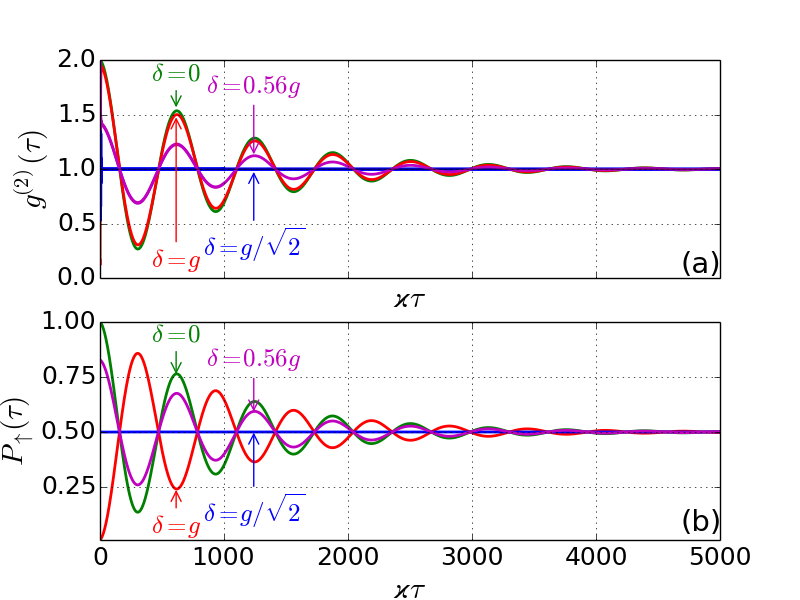}
  \caption{(a) The photon correlator $g^{(2)}(\tau)$ at long timescales calculated for the same parameters as in Fig.~\ref{fig:g2t}.
  (b) Spin-up population $\tilde{P}_\u(\tau)$.
}
  \label{fig:Pud}
\end{figure}

The same calculation is shown in Fig.~\ref{fig:Pud}(a) in linear scale: $g^{(2)}(\tau)$ very
rapidly reaches the value $g^{(2)}_{IT}$ during the short timescales, and then oscillates with the period $T_L=2\pi/\Omega_L$. For comparison the time dependence of $\tilde{P}_{\u}(\tau)$ is shown in Fig.~\ref{fig:Pud}(b) for the same parameters. We note that the phase of the oscillations of $\tilde{P}_{\u}(\tau)$ reverses with the detuning, while the phase of oscillations in $g^{(2)}(\tau)$ is constant. This can be easily understood as follows: for example, let us consider the detuning when $T_\u\gg T_\d$. In this case the detection of a photon at time $t$ projects the electron in the spin-up state. The detection of the second photon again requires the spin to be in the same state, which happens after the delays $\tau=k T_L$ with $k=1,2,\ldots$. Therefore the phase of the oscillations in $g^{(2)}(\tau)$ is always the same, and this conclusion is not changed for any relation between $T_\u$ and $T_\d$. However the amplitude of the oscillations depends on the detuning, and in the case of $T_\u=T_\d$ the oscillations are absent, since $g^{(2)}(\tau)=g^{(2)}_{IT}=1$, see the blue curves in Figs.~\ref{fig:Pud}(a) and (b).

Importantly, and in contrast with standard spin noise spectroscopy, we stress that a single photon detection is sufficient to fully initialize a spin at $\tilde{S}_z(0)=-\frac{1}{2}$ or $+\frac{1}{2}$, leading to a maximally-polarized single spin precession. In the event that $\Omega_L \tau_s \gg 1$, this also means that the single spin can experience a coherent Larmor precession, described by a pure state evolving at the surface of the Bloch sphere, before decoherence occurs. Such a heralded generation of coherent spin precession can be directly used in several quantum optics protocols, where the manipulation of coherent spin superpositions is a crucial element \cite{Hu2008a,Hu2008,Lindner2009}.

We also point out that the overall time dependence of the second order photon correlator can be presented as
\begin{equation}
  g^{(2)}(\tau)=
  2\tilde{P}_\u^2(0)
\left[g^{(2)}_\u(\tau)-1\right]
+4 \tilde{S}_z(0) \tilde{S}_z(\tau)
+1
.
\label{eq:Call}
\end{equation}
This is the main result of this paper. This expression is plotted in Fig.~\ref{fig:g2t} with the dashed lines, in agreement with the numerical results.
The first term on the right hand side of Eq.~\eqref{eq:Call} describes the time dependence of $g^{(2)}(\tau)$ at short timescales, as it oscillates and changes from $g^{(2)}(0)$ to $g^{(2)}_{{IT}}$. The second term describes the time evolution on the long timescales, when $g^{(2)}(\tau)$ changes from $g^{(2)}_{{IT}}$ to unity oscillating at the Larmor frequency. One can see that the dynamics of $g^{(2)}(\tau)$ is completely described only by the time dependence of $g^{(2)}_\u(\tau)$, describing the photon-polariton correlations, and the value of $\tilde P_\u(0)$, describing the quantum back-action induced by the photon detection on the spin system. Indeed the latter governs the amplitudes of both the short and the long timescale fluctuations, because $\tilde{S}_z(0)=\tilde P_\u(0)-1/2$. 

Finally, we also stress that Eq. \ref{eq:Call} will be valid as well in the weak-coupling regime of cavity-QED, where giant spin-dependent signals can also be obtained \cite{Arnold2015}.  While the expression of $g^{(2)}_\u(\tau)$ would differ in this regime, leading to a different evolution at short timescales, the effect of single-spin projection and the corresponding Larmor precession at longer timescales will be strictly identical to what we have described. This feature is especially important as it ensures a possible experimental demonstration of the effect in very realistic devices. 

\section{Transmittance noise spectrum}

\label{sec:spectrum}

The transmittance fluctuations can be conveniently studied by means of spin noise spectroscopy~\cite{Oestreich-review,Zapasskii:13}. The spectrum of fluctuations is defined as
\begin{equation}
  C(\Omega)=\int\limits_{-\infty}^\infty C(\tau)\e^{\i\Omega\tau}{\rm d}\tau,
  \label{eq:def_spectrum}
\end{equation}
where the correlator $C(\tau)=g^{(2)}(\tau)-1$ [Eq.~\eqref{eq:Cg2}] should be used to avoid a delta function in the spectrum at zero frequency. It can be calculated analytically using Eqs.~\eqref{eq:g2u} and~\eqref{eq:g2_short}, however the result is too cumbersome to be presented here.

\begin{figure*}
  \includegraphics[width=0.49\linewidth]{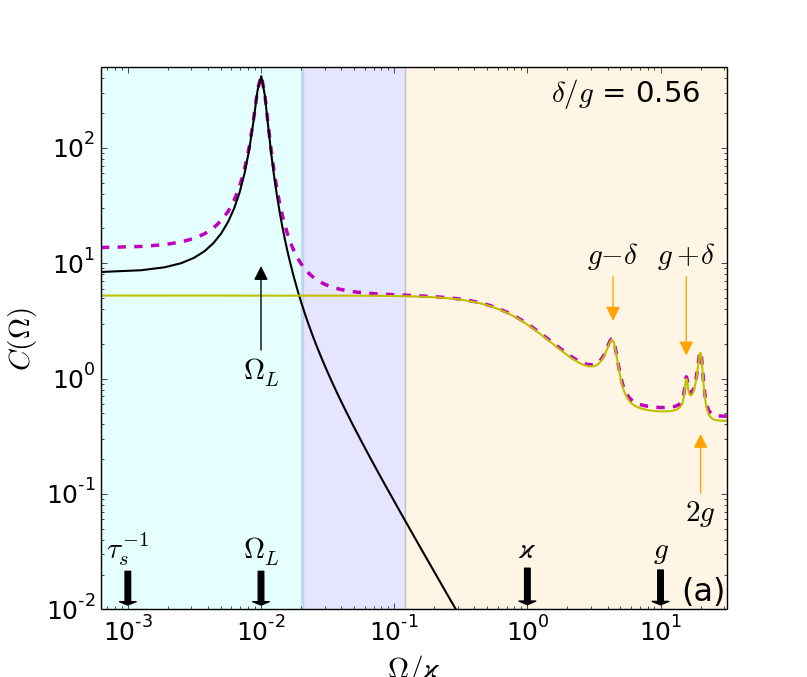}
  \includegraphics[width=0.49\linewidth]{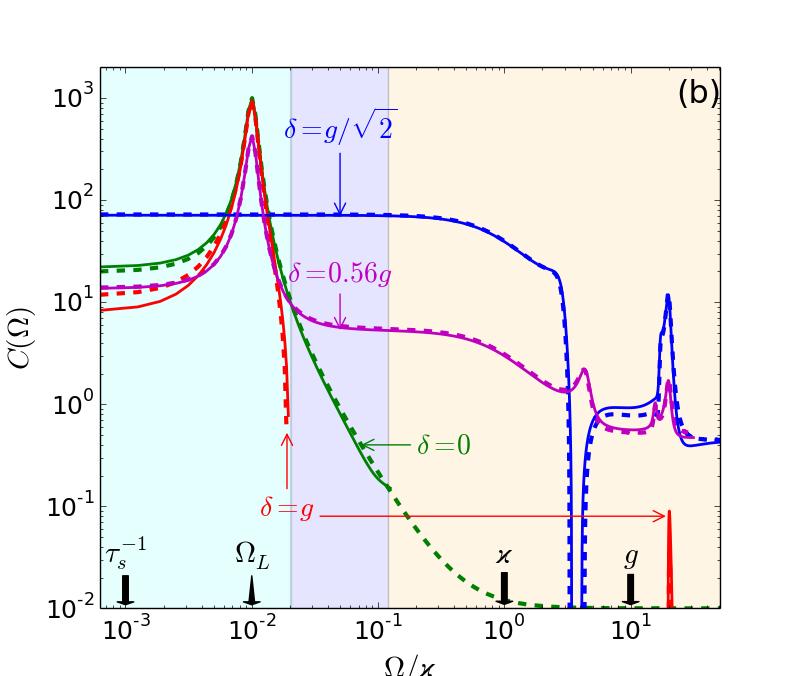}
  \caption{(a) Spectrum of transmittance fluctuations at detuning $\delta=0.56 g$ (dashed magenta line) calculated after Eqs.~\eqref{eq:Call} and~\eqref{eq:def_spectrum}. The photonic part of the spectrum is shown by the yellow line, and the magnetic part by the black solid line.
(b) Spectra of transmittance fluctuations for the detunings indicated in the legend. The dashed lines represent the analytic calculation and the solid lines are calculated numerically. The parameters of the calculation are the same as in Fig.~\ref{fig:g2t}. The background colors indicate frequency ranges that correspond to the timescales separation as in Fig.~\ref{fig:g2t}.}
  \label{fig:spectra}
\end{figure*}

The exemplary spectrum for the detuning $\delta=0.56 g$ is shown in Fig.~\ref{fig:spectra}(a). The spectrum is an even function of $\Omega$, so only the positive frequencies range is shown. At small frequencies $\Omega\ll\varkappa$ the spectrum is described by a pair of Lorentzians at frequencies $\Omega=\pm\Omega_L$: 
\begin{equation}
  C(\Omega)=4 \tilde{S}_z^2(0) \left[\frac{\tau_s}{1+(\Omega-\Omega_L)^2\tau_s^2}+\frac{\tau_s}{1+(\Omega+\Omega_L)^2\tau_s^2}\right],
\end{equation}
which corresponds to the Larmor precession of electron spin. This contribution can be called the magnetic part of the spectrum. It is shown by the black curve in Fig.~\ref{fig:spectra}(a).

The rest of the spectrum [yellow line in Fig.~\ref{fig:spectra}(a)] originates purely from photon-polariton correlations, and can be called the photonic part of the spectrum. This part consists of four additional peaks at the frequencies $\Omega=0$, $2g$ and $\left|g\pm\delta\right|$. 
The half widths at the half maximum of the first two peaks are $\varkappa$, and those of the two others are $\varkappa/2$, therefore they are all determined by the lifetime of the photon inside the cavity. Note that
the peak at $\Omega=0$ is hindered by the spin precession peak.

The shape of the spectrum and, in particular, the amplitudes of the peaks strongly depend on the light frequency, as shown in Fig.~\ref{fig:spectra}(b) for various detunings. Note that the numerical calculations coincide with the analytical expressions within numerical accuracy. At zero detuning, $\delta=0$, (green curve) the photonic part of the spectrum is absent, because $g^{(2)}(0)=g^{(2)}_{IT}$, and the spectrum consists of a single Lorentzian peak at the Larmor frequency. By contrast at $\delta=g/\sqrt{2}$ (blue curve) the spin part vanishes: $T_\u=T_\d$ and thus the spin remains unpolarized after a photon detection. Interestingly the peak at $\Omega=g-\delta$ in this case has a negative amplitude, and appears as a dip in the spectrum. This is the consequence of the so-called weak positivity property of the noise spectrum~\cite{Bednorz_PRL}. This effect is most pronounced at the detuning $\delta=g$ (red curve), where nearly the whole photonic part of the spectrum is negative. Indeed the area under the spectrum is proportional to $\mathcal C(0)=g^{(2)}(0)-1$, so in case of antibunching the area under the transmission noise spectrum is negative.

The peak at the frequency $\Omega=2g$ corresponds to the Rabi oscillations between the lowest polariton states with $m=1$. These oscillations correspond to the case when one photon is extracted from one of the secondary polariton states with $m=2$. Therefore this peak has the maximum amplitude at the detunings $\delta=\pm g/\sqrt{2}$, when one of such states is resonantly excited. The peak at $\Omega=0$ corresponds to the monoexponential loss of coherence of the polariton states due to the escape of the cavity photons. Finally, the peaks at the frequencies $\Omega\approx\left|g\pm\delta\right|$ are most pronounced at the detunings $\delta=\pm g$, when the system behaves as a two level atom. They correspond to the Rabi oscillations between the ground and excited states~\cite{carmichael1985photon}. As one can see, measuring the spectrum of transmission fluctuations allows determining the strength of the measurement back-action, but also all the relevant quantities governing the device dynamics: polariton frequencies, decay rates of the excited states, QD-cavity detuning, spin relaxation time, and Larmor precession frequency.




\section{Conclusion}
\label{sec:conclusion}

In this work we have studied the intensity fluctuations of light passing through a charged quantum dot cavity-QED device in the strong coupling regime. Depending on the electron spin orientation, the incoming photons can be completely transmitted or completely reflected from the structure. Giant spin-induced fluctuations are thus obtained, characterized by a second-order correlation function $g^{(2)}(\tau)$, which can be analytically calculated in the low intensity regime. At short timescales, this function is found to reflect the damped Rabi oscillations between polariton states, as well as the optical nonlinearity induced by photon-polariton interactions. At long timescales, it reflects the Larmor spin precession in an external magnetic field, and its progressive relaxation. Interestingly, it also reflects the quantum back-action induced by a single photon detection on the spin system. At intermediate timescales, in particular, the second-order correlation is found to provide a quantitative measure of the back-action strength.

In addition, the spectrum of these fluctuations could be analytically calculated, displaying a number of peaks whose positions and widths are directly given by the system parameters: vacuum Rabi frequency, spectral detuning, polariton decay rate, Larmor precession frequency, and spin relaxation time. 

One direct implication of these results is that perfect back-action induced by photon detection can be used to trigger and monitor a coherent Larmor spin precession, providing a very useful tool for various quantum applications. The possibility to control a coherent spin oscillation, and probe it with successively incoming photons, is indeed at the heart of many propositions for spin-photon entanglement \cite{Hu2008a}, delayed-photon entanglement \cite{Hu2008}, and multiphotonic cluster state generation \cite{Lindner2009}.

Finally, we stress that all the analytical results obtained in the low intensity regime are found in agreement with numerical calculations. The numerical approach allows investigating the system at arbitrary large incoming light power; in such a configuration the spin evolution is modified by multiple back-action events induced by numerous incoming photons. This paves the way towards the possible demonstration of the quantum Zeno effect \cite{Itano1990}, where the spin evolution is frozen by the fast repetition of strong projective measurements.

\begin{acknowledgements}

We thank M. M. Glazov for stimulating and fruitful discussions. This work was partially supported by the Russian Foundation for Basic Research (grant No. 17-52-16020), RF President Grant SP-643.2015.5 and by the French Agence Nationale pour la Recherche (SPIQE: ANR-14-CE32-0012).

\end{acknowledgements}


%

\end{document}